\documentclass[prd,floats,preprint,superscriptaddress,eqsecnum,floatfix,nofootinbib,12pt]{revtex4}
\pdfoutput=1
\usepackage{amsmath}
\usepackage{graphics, graphicx}
\usepackage{epsfig}
\usepackage{ulem}
\usepackage{color}
\usepackage{booktabs}
\usepackage{bm}

\newcommand\hcancel[2][black]{\setbox0=\hbox{$#2$}%
\rlap{\raisebox{.45\ht0}{\textcolor{#1}{\rule{\wd0}{1pt}}}}#2}

\def\1p{}
\def\be{\begin{equation}}
\def\ee{\end{equation}}
\def\beq{\begin{eqnarray}}
\def\eeq{\end{eqnarray}}
\def\p0{\phi_0}
\def\z0{\zeta_0}

\def\3G{^3{\cal G}}

\def\vx{{\vec x}}

\def\pfp{p}

\def\ol2{\frac{1}{\ell^2}}

\def\St0{\Sigma(t_0)}

\def\HP{$(I,\Psi) \ $}

\def\Dl0{D_{\rm loc}}

\def\Daoi{D^{\ge 1}}

\def\Eo{E^{\ge 1}}
\def\Ai{$A$}
\def\Bi{$B$}
\def\av{\vec\alpha}
\def\ab{\bm\alpha}

\def\Kb{{\bf K}}
\def\bb{\bm\beta}
\def\Gm{\Gamma}
\def\Pss{{\sf P}}
\def\uf{}
\def\qf{}
\def\zf{}
\def\zzf{}
\def\rf{}

\def\nnf{}

\def\yf{}

\def\wwf{}
\def\wzf{}

\def\qf{}
\def\qqf{}

\newcommand{\ttle}[1]{{\it #1}}

\begin{document}

\vspace{1cm}

\title{One Bubble to Rule Them All}

\author{James  Hartle}
\affiliation{Department of Physics, University of California, Santa Barbara,  93106, USA}
\affiliation{Santa Fe Institute,  Santa Fe, NM } 
\author{Thomas Hertog}
\affiliation{Institute for Theoretical Physics, KU Leuven, 3001 Leuven, Belgium}

\bibliographystyle{unsrt}

\vspace{1cm}

\begin{abstract}

We apply the principles of quantum mechanics and quantum cosmology to predict probabilities for our {\zf local} observations {\zf of} a universe undergoing false vacuum eternal inflation. At a {\zf sufficiently} fine-grained level, histories of the universe {\zf describe} a mosaic of bubble universes separated by inflationary regions. We show that predictions for local observations can be obtained directly from sets of {\zf much coarser grained histories which only follow} a single bubble. {\zf These} coarse-grained histories contain neither information about our unobervable location nor about the {\zf unobservable} large-scale structure outside our own bubble. Applied to a landscape of false vacua in the no-boundary state we predict {\zzf our local universe emerged from the dominant decay channel of the lowest energy false vacuum}. We compare and contrast {\zf this} framework for prediction based on quantum cosmology with traditional approaches to the measure problem {\wwf in cosmology}.

\end{abstract}

\vskip.8in


\maketitle

\tableofcontents


\section{Introduction} 
\label{intro2}

The string theory landscape \cite{Landscape} is believed to contain a vast ensemble of false vacua including some with a positive effective cosmological constant and the low energy effective field theory of the Standard Model. But the landscape by itself does not predict why we are in one vacuum rather than in some other. For that one has to turn to cosmology. 

A quantum theory of cosmology that consists of a model of the quantum state of the universe combined with the structure of the landscape potential {\it predicts} a prior that specifies predictions for our observations \cite{HHH10b,Her13}. This prior -- or measure -- takes the form of probability distributions for cosmological observables. Observables with distributions that are sharply peaked around specific values are predicted with high accuracy by the theory and can be used to test it against observations.

{\rf This paper gives} a quantum cosmological derivation of probabilities for our observations in a landscape where our universe can undergo false vacuum eternal inflation. {\wzf Our discussion is a generalisation, and to some extent a justification, of our work in \cite{HHH10b}.} We show in particular that predictions for local observations can be obtained directly from {\rf suitable sets of alternative} coarse-grained histories of the universe. The key roles played by coarse graining and by utilizing all available symmetries sharply distinguishes our derivation of these predictions from the traditional approach to the measure problem in eternal inflation (TEI).

Quantum mechanics allows descriptions of any system at different levels of coarse graining. For example, at a very coarse grained level, the Moon can be described by the position of its center of mass. At a very-fine-grained level it could be described by the position and quantum state of every atom it contains. Predictions for our observations of the Moon's orbit could be derived from either, but the coarse graining that focuses on the center of mass that we observe is simpler and more manageable. {\rf In the language of Feynman's path integral the individual  paths are the finest grained description  of a system. Coarse- grained paths that follow only certain features are bundles of the fine-grained paths that have  those features.} {\zzf Coarse graining can be carried out directly at the level of amplitudes by summing over the paths in the bundle, without summing probabilities of a finer grained description.} Of course, a measure is involved in this coarse graining too, but it is the measure that defines the path integral, essentially $dpdq/\hbar$. 

In {\zzf eternal inflation} a fine-grained description of a particular {\rf four-dimensional} history of the universe consists of a complicated {\rf mosaic} of various kinds of {\zzf pocket universes} separated by inflationary regions \cite{Star86,EI,HHH10a}. {\rf TEI  aims at }deriving predictions for our observations in a universe of this kind {\rf starting from} a fine-grained description of one of its histories. Specifically, in TEI the relative likelihood of two different observations is argued to be proportional to the number of observations of each kind in a typical fine-grained history of the universe. To compute this one must first regularize the divergences that arise if one extends the fine-grained description into the far future, leading to an infinite number of pocket universes and therefore to ambiguities. A procedure for regulating the infinities is called a measure in TEI and is usually thought of as a supplement to the theory.  A TEI measure typically consists of specifying a geometric cutoff in the form of a spacelike three-surface beyond which one no longer counts instances of observations (for examples of TEI measures see e.g. \cite{TEI}). However the resulting predictions turn out to depend on the choice of cutoff \cite{ProblemsTEI}, and {\wwf to be} in conflict with basic physical intuition \cite{EOT}.

Our observations of our universe are very, very coarse grained. They are essentially limited to what goes on in our Hubble volume. In the fine-grained description of {\zzf false vacuum eternal inflation} that is used in TEI our Hubble volume is but one of an infinite number of other Hubble volumes on a constant density surface inside a particular bubble that is but one of an infinite number of bubbles in the larger universe. Hence it would be more natural to use a much coarser grained description to calculate the probabilities for our observations than what is done in TEI. {\zzf This requires a quantum framework for prediction, however, which involves a sum over histories that restores the probabilistic symmetry between different Hubble volumes. As we will see the basic principles of quantum mechanics can then be used to} calculate probabilities for our coarse-grained observations simply and directly, without using a fine-grained TEI kind of description, and therefore without a measure beyond that supplied by the quantum state. 

In our analysis we assume that the universe is a quantum mechanical system with a quantum state $\Psi$ and a dynamics specified by an action $I$. Further we work in a low energy approximation in which the fields are  four-dimensional spacetime geometries coupled to matter fields. As an illustration in this Introduction we model the matter in terms of a single scalar field with a potential that has one false vacuum with two decay channels to two different true vacua $A$ and $B$. We assume that the slow roll conditions for inflation hold in the neighborhood of $A$ and $B$, and that the slow roll evolution towards both true vacua yields a statistically different pattern of CMB fluctuations that provides an observational discriminant between $A$ and $B$ \cite{ObsLandscape}. 

{\zzf At a sufficiently fine-grained level, the histories of the universe in this model describe a mosaic of bubble universes separated by inflationary regions. The quantum state together with the different decay rates specify quantum mechanical {\it bottom-up probabilities} for the individual members of decoherent sets of histories of the universe at different levels of coarse-graining. However we are not so much interested in the bottom-up probabilities for what occurs.
Rather, we are interested in the {\it top-down probabilities} following from \HP that {\it we} observe the properties of a bubble of type $A$ or $B$ (assuming that we don't live in the false vacuum\footnote{{\wzf This amounts either to a typicality assumption that we are not Boltzmann brains \cite{HS07} or to a requirement on the potential that the dominant decay rate of all positive false vacua is larger than the nucleation rates of a Boltzmann brain consisting of data $D$ (see e.g.~\cite{TEI}).}}. The history that is most probable to occur is not necessarily the most probable to be observed \cite{HH06}. This is because top-down probabilities are bottom-up probabilities conditioned on there being at least one instance of our observational situation $D$ \cite{HH09,HHH10b}.} 

To evaluate top-down probabilities therefore requires a model of an observer. In quantum cosmology observers are modeled as physical quantum systems within the universe, describable in physical terms and subject to quantum mechanical laws. {\rf As physical systems particular} observers have a very, very small probability  to have evolved in any one Hubble volume. But in the large universes implied by false vacuum eternal inflation inflation the probability may be significant that the same system has evolved in many different Hubble volumes in many different bubbles \cite{HS09,HH09}. We are one of these instances, in one Hubble volume, in one bubble, somewhere in the vastness of the spacetime of an eternally inflating universe. 
The theory \HP does not specify which of these instances is us. Rather we calculate the top-down probabilities of our observations assuming that we are equally likely to be any of the instances. We adopt a crude model of observers in terms of systems {\rf described by data} $D$ which either exist or do not exist in any Hubble volume, with probabilities $p_E(D)$ and $1-p_E(D)$ that may depend on the type of bubble \cite{HS09,HH09}. That is a highly simplified model of an observer but it is a fundamental improvement over TEI where the observer is regarded as a classical system outside the universe -- certain to exist wherever it can.

Exploiting the {\it probabilistic} symmetry between bubbles at different locations implied by the false vacuum background we shall calculate the top-down probabilities for our local observations using a coarse-grained description that follows only one bubble -- ours -- and ignores everything outside. When the nucleation rate for both kinds of bubbles is low enough that bubble collisions can be neglected\footnote{When the nucleation rate is higher we exhibit a systematic way of improving this result to include bubble collisions.}, we find for the probabilities $p(WOA)$ and $p(WOB)$ that we observe the CMB properties of $A$ or $B$: 
\be
\label{obs1}
p(WOA)=\frac{p_A}{p_A+p_B} ,  \quad p(WOB)=\frac{p_B}{p_A+p_B},
\ee
where $p_A$ is the probability of a coarse-grained history {\rf where the one bubble is} of type $A$ in the quantum state $\Psi$, and similarly for $B$. Assuming that the reheating surfaces are infinite in either kind of bubble the predictions do not depend on the probabilities $p_E(D)$ provided that they are not zero. 

{\zzf In the following sections we {\wwf first develop our general framework and then apply it} in the context of} more complicated landscape potentials with several false vacua at different values of the potential. In the no-boundary quantum state \cite{HH83}, {\zzf we find that the dominant contribution to top-down probabilities comes from the coarse-grained history that follows the lowest action bubble describing the dominant decay of the lowest false vacuum. This generalizes our previous result \cite{HHH10b,Her13} that top-down probabilities are dominated by the lowest exit from eternal inflation in the no-boundary state.

Actually, we will see that the probabilities of the coarse-grained histories that are relevant for top-down probabilities can be computed directly from the no-boundary wave function.} The saddle points are either of the Hawking-Moss type \cite{HM} or a no-boundary version of the Coleman-De Luccia instanton \cite{CDL} in which a bubble of true vacuum emerges together with the false vacuum background. The fact that these {\wwf saddle points yield the amplitude of} coarse-grained histories in our work leads us to conclude they contain neither information about our unobervable location nor about the unobservable large-scale structure outside our own bubble. {\zzf This in turn provides a novel interpretation of the nature of the no boundary condition, as a sum over all possible past histories.}

{\zzf The paper is organized as follows: Section \ref{YGmodel} contains a toy model in which coarse graining can be illustrated explicitly. Section \ref{EI} introduces a simple model of false vacuum eternal inflation. In Section \ref{cgdesc} we discuss the ensemble of fine-grained quasiclassical bubble histories in this model and their bottom-up probabilities. Section \ref{cgobs} is concerned with the set of coarse-grained histories following a single bubble only that are appropriate for the prediction of top-down probabilities of local observations. In Section \ref{probnucl} we use these to evaluate the top-down probabilities of CMB related observables in the no-boundary state in a toy model landscape with a number of different false vacua. Finally in Section \ref{compare} we compare and contrast this framework for prediction based on quantum cosmology with traditional approaches to the measure problem.}

\section{A Simple Box Model}
\label{YGmodel} 
\begin{figure}[t]
\includegraphics[width=5in]{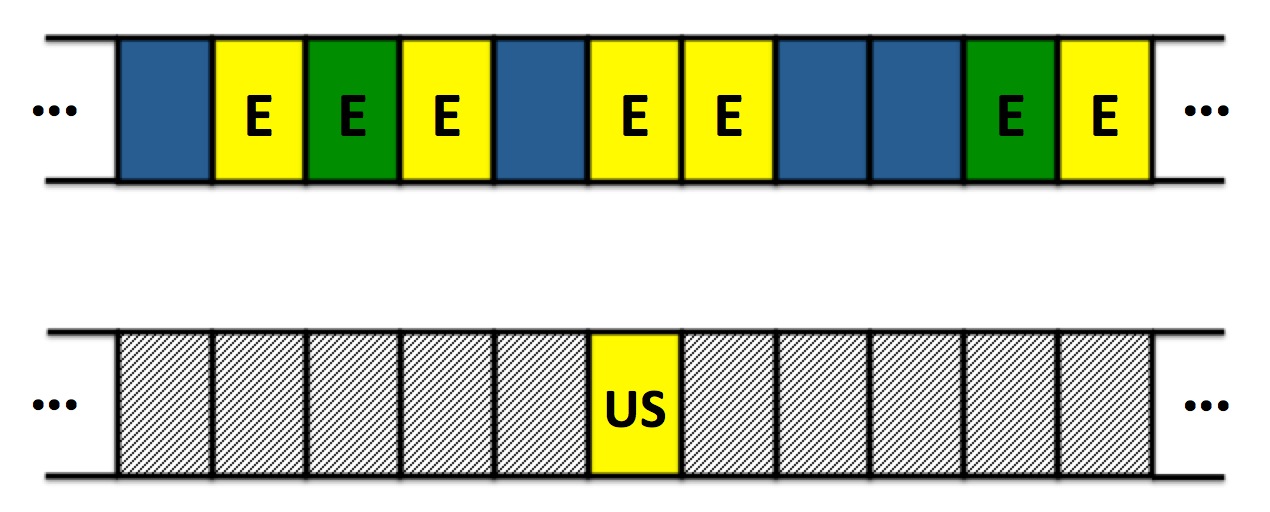}\hfill 
\caption{Fine and coarse-grained histories of  a box model discussed in the text. The boxes model bubbles of true vacuum. The colors yellow and green model observable properties like the CMB inside of two kinds of true vacuum bubbles $A$ or $B$. Blue means the Hubble volume is in false vacuum $F$. An `$E$' means that {\qqf at least one observer is in the box observing its color (denoted by $\Eo$ in the text)}. No `$E$' means no observer. {\qqf The top history is fine grained with a color  in every box and  $E$ in every true vacuum box and no $E$ in every false vacuum box.} The bottom history is coarse grained. The possibilities have been summed over for every box except the one we are in. The result does not depend on {\wwf where we are}. The coarse-grained boxes are gray.  Coarse graining  enables a straightforward, direct,  calculation of the {\wwf top-down} probability that we see one color or the other without calculating probabilities for detailed fine-grained structure.}
\label{YGfig}
\end{figure}

We illustrate how coarse graining enables the calculation of predictions for local observations employing a  simple box model universe of the kind used in \cite{HH15b,HS07,HS09}. A box model universe at one time consists of $N$ boxes as illustrated in Figure \ref{YGfig}.  To distinguish the boxes we  label them with an index $i=1,2,\dots N$.  The model  universe has a quantum state $\Psi$ that predicts probabilities for what goes on in each box as follows: Each box can be of one of three kinds $K=A, B, F$ with probabilities $p_A, \ p_B, \ p_F$ --- the sum of these adding to unity, $\sum_K p_K =1$.  Boxes $A$ and $B$ model different kinds of bubbles of true vacuum distinguished by  different  probabilistic predictions for observables like the CMB. The alternative $F$ models a false vacuum. 

Observers  like us are physical systems within the universe. We have a small probability to have evolved in any one box but, in a very large universe, a significant probability to be replicated in other boxes. {\qf What we know is that there exists at least one copy --- us --- in one box. We denote by $p(\Eo|K)$  the probability that there is at least one observing system like us in a box of type $K$. 
{Bubbles of true vacuum} contain spacelike surfaces with an infinite number of Hubble volumes. Then, even though the probability of an observer like us in any one Hubble volume  is very small, the probability that there is an observer like us in any bubble is very near unity.   In this box model we therefore take for the probabilities of at least one observer in a box
\be
\label{pE1}
p(\Eo|A)=p(\Eo|B) =1, \quad  p(\Eo|F)=0 .
\ee
Thus there is always a copy of us in any box of type $A$ or $B$ and no copy in a box of type $F$. {\yf In the Appendix the model is generalized to deal with arbitrary probabilities for $p(\Eo|K)$.}

In this model universe all the boxes are probabilistically identical: neither $p_K$ nor $p_E(K)$  depend  on $i$ labeling the specific box. In this sense this model has a discrete probabilistic translation  symmetry in $i$ traceable to the quantum state. This symmetry is broken in any particular history of the universe specified by the alternatives $(A,B,F)$ for each box, and whether there exists an observer in it or not $(E, \bar E)$ (cf. top in Figure \ref{YGfig}). The probability for a fine-grained history having $N_A$ $A$-boxes and $N_B$ $B$-boxes is proportional to 
\be
\label{fgprobs}                       
(p_A)^{N_A}(p_B)^{N_B}(1-p_A-p_B)^{N-N_A-N_B} . 
\ee

The probabilities \eqref{fgprobs} depend crucially on $N$. They tend to zero in a universe with a large number of boxes. In the infinite $N$ limit --- the analog  of eternal inflation --- the probabilities for these fine-grained histories are not well defined. 
Well defined results could be obtained by  imposing a cut-off procedure or `measure'. But then the results will generally depend on the procedure. In the following we will show that a cut-off is not  necessary to get well defined probabilities for our observations. 

Our observations are limited to one box, but don't tell us which.  All the boxes are probabilistically the same with respect to {\wwf local} observables, and the label $i$ is not observable. {\wwf Hence} what we observe depends {\wwf only} on what's inside the box, not what's outside.  {\wwf This means that the} detailed large scale structure outside the box is irrelevant for our local observations. Finite probabilities for observations can thus be obtained by coarse graining { over everything outside}. That is, they can be obtained by summing \eqref{fgprobs} over all the alternatives outside the box we are in.  That means summing\footnote{For an explicit calculation of the sum in a similar box model see the Appendix of \cite{HS09}.} the probabilities \eqref{fgprobs} over the alternatives $(A,B,F)$ and $(E,\bar E)$ giving a factor of unity for every box but the one we are in  (cf. bottom in Figure \ref{YGfig}).

The box we occupy must be one of the true vacua $A$ or $B$.  We denote the probabilities  that we observe the alternatives $CMB_A$ or $CMB_B$ by $p(WOA)$ and $p(WOB)$. The probability  that $WOA$ is the probability that our box is of type $A$ given that there is at least one observer in the box  i.e. $p(A|\Eo)$. Similarly for $WOB$. The probabilities $p(WOA)$ and $p(WOB)$ are examples of  top-down or first person probabilities that specify predictions for our observations, whereas the probabilities in  \eqref{fgprobs} are bottom-up or  third person  probabilities for what occurs in the universe\footnote{For a more extended discussion than is reasonable to give here see \cite{HH15b} .}. 
  Since both $A$ and $B$ imply the existence of an observer \eqref{pE1} the joint probability $p(A,E) $ is $p_A$. But we could also be in a type $B$ box for which $p(B,E)=p_B$.
 The total probability for $\Eo$ is therefore $p(\Eo)=p_A +p_B$ and the conditional probabilities  $p(WOA)\equiv p(A|\Eo)$ and  $p(WOB)\equiv p(B|\Eo)$ are [cf \eqref{obs1}] }
\be
\label{obs}
p(WOA)=\frac{p_A}{p_A+p_B} ,  \quad p(WOB)=\frac{p_B}{p_A+p_B},
\ee
which are correctly normalized. 
These probabilities for $WOA$  and $WOB$ for our observations are independent of $N$ and remain well defined in the infinite $N$ limit. The fine grained probabilities \eqref{fgprobs} depend on $N$ and are ambiguous in that limit. 

With \eqref{obs} we have succeeded in calculating coarse grained probabilities for our local observations directly, without explicitly considering what large scale structure the universe might have.  One has only to compare the top and bottom parts of Figure \ref{YGfig} and the probabilities \eqref{obs} and \eqref{fgprobs} to see that both the description of the coarse grained histories and the probabilities for the results are much simpler - and less ambiguous - than for the fine grained ones. The simplifying assumption in this model is a symmetry arising from the wave function, implying that all boxes are probabilistically the same. This simplicity of the quantum state is not evident in any one fine-grained history because any history is a particular instance of a large number of random events and correspondingly complex to describe. The simplicity of the state is rather manifest in the ensemble of possible fine-grained histories or in coarse grainings that are sums over these. 

To summarize, in infinite or just very large universes focusing on our observations motivates coarse grainings that directly lead to well defined  probabilities for those observations even when the probabilities for the very fine-grained structure outside our Hubble volume may not be well defined. 

\section{False Vacuum Eternal Inflation}
\label{EI}

We now turn to the quantum cosmology of false vacuum eternal inflation.  We consider four-dimensional Einstein gravity coupled to a single scalar field  $\phi$ moving in a positive potential as a model of the dynamics $(I)$. 
We take the potential to have a false vacuum $F$ with two quantum mechanical decay channels to two different true vacua, $A$ and $B$  where the potential vanishes. {\uf  Figure \ref{pot} shows an example.} {\uf The classical equations following from this theory have an eternally inflating de Sitter solution {\nnf with an effective cosmological constant given by} the value of the potential in the false vacuum}. {\uf Quantum mechanically this solution decays through the nucleation of bubbles of true vacuum. The geometry inside these bubbles is that of an open universe which expands in the de Sitter background \cite{CDL}.} We will allow for different decay rates of the false vacuum to \Ai\ and \Bi . We will assume that the nucleation rates of both kinds of bubbles are small so that we can ignore collisions between bubbles. {\qqf We return to models with higher nucleation probabilities in Section \ref{probnucl} below. }

We will further assume that the potential towards \Ai\ or \Bi \ has flat patches where the slow roll conditions hold so that {\nnf the open universes inside bubbles undergo} a period of inflation during which the scalar slowly rolls down to one of the true vacua, before the universe reheats and standard cosmological evolution ensues. Finally we assume the potential is such that CMB related observables such as the spectral tilt etc enable observers inside one of the bubbles to determine whether they live in $A$ and $B$.

For {\nnf  most of} this paper we will adopt the no-boundary wave function (NBWF) as a model of the quantum state $\Psi$ of the universe \cite{HH83}. The NBWF is defined on the superspace of three metrics $[h_{ij}(\vec x)]$ and spatial scalar field configurations $[\chi(\vx)]$ on a closed spacelike three-surface $\Sigma$. 
In the semiclassical approximation it is given by\footnote{We use Planck units where $\hbar=c=8\pi G=1$.} {e.g. \cite{HHH08a,HHH08b}) 
\begin{equation}
\Psi[h_{ij},\chi] \propto  \exp(-I[h_{ij},\chi]) = \exp(-I_R[h_{ij},\chi] +i S[h_{ij},\chi])
\label{semiclass}
\end{equation}
where $I_R[h_{ij},\chi]$ and $-S[h_{ij},\chi]$ are the real and imaginary parts of the Euclidean action $I$ of a compact {\it regular} saddle point history $(g_{\mu \nu},\phi)$ that matches the real boundary data $(h_{ij},\chi)$ on its only boundary $\Sigma$. The Euclidean action of the model universe we consider is given by
\begin{equation}
I [g_{\mu \nu},\phi] = \int_M d^4 x (g)^{1/2}\left[- \frac{1}{2} R + \frac{1}{2} (\nabla \phi)^2 +V(\phi) \right] + \int_{\partial M} d^3 x (h)^{1/2}K
\label{Eact}
\end{equation}

\begin{figure}[t]
\includegraphics[width=4in]{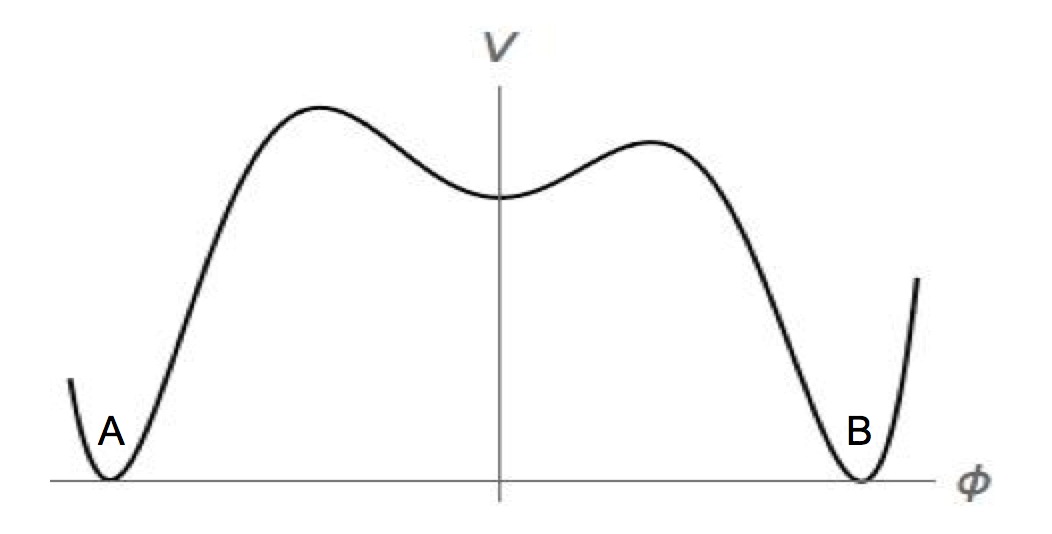}\hfill 
\caption{A potential with one false vacuum F and two true vacua A and B. The false vacuum is separated from both true vacua by a potential barrier and a relatively flat patch where the slow roll conditions for inflation are satisfied. The different shape of the barriers and of the potential in the two slow roll regimes leading to the true vacua results in different false vacuum decay rates and different predictions for CMB related observables in universes ending up in $A$ or $B$.}
\label{pot}
\end{figure}

In regions of superspace where $S$ varies rapidly compared to $I_R$ the wave function takes a WKB form and predicts 
Lorentzian, classical evolution. The NBWF has the striking property that it selects classical histories in which the universe undergoes some amount of scalar field driven inflation at early times \cite{HHH08a,HHH08b}. Intuitively this is because only universes with sufficiently small gradients initially can be obtained from compact regular saddle points. 

In particular, in the single field potential illustrated in Fig \ref{pot} the NBWF selects a one-parameter set of inflationary backgrounds {\nnf with different starting values of the scalar field.} It is convenient to label the individual histories in the classical ensemble predicted by the NBWF by the absolute value $\phi_0$ of the field at the `South Pole' (SP) of the corresponding saddle point, where the scale factor vanishes \cite{HHH08b}. It turns out that $\phi_0$ is approximately equal to the {\nnf initial value of the inflaton} in the Lorentzian history. {\nnf There are two qualitatively different kinds of inflationary histories in the false vacuum model of Fig \ref{pot}.} First there are saddle points with $\phi_0$ on the potential slope around \Ai \ or \Bi\ {\nnf where the slow roll conditions hold}. These correspond to closed universes with a period of slow roll inflation, and no eternal inflation. Second there is the saddle point with $\phi_0=0$ which corresponds to the false vacuum background. 
 
The individual histories in the classical ensemble are given by the integral curves of $S$. Their tree-level probabilities in the no-boundary state are proportional to $\exp[-2 I_R(h_{ij},\phi)]$ {\yf evaluated on these curves}.  To a good approximation one finds \cite{HHH08a}:
\be
\exp[-2 I_R] \approx \exp \left[\frac{24\pi^2}{V(\phi_0)}\right] ,
\label{bu}
\ee
which is constant along the classical histories as a consequence of the Wheeler-DeWitt (WDW) equation for $\Psi$. 

{\yf Evidently this ensemble of classical histories coarse-grains over all quantum events. In particular the classical ensemble says nothing about the quantum nucleation of Coleman-DeLuccia (CDL) or Hawking-Moss (HM) bubbles {\nnf in the false vacuum background,} within which the universe decays to one of the true vacua.} Put differently, the classical ensemble is far too coarse-grained to describe the possibility that we find ourselves in a bubble universe. To take in account bubbles we must fine-grain the classical ensemble and consider the ensemble of {\it quasiclassical histories.} This is the subject of the next Section.

\section{Quasiclassical Histories of the Universe}
\label{cgdesc}

\subsection{Coarse Graining and Classicality}
\label{cgcl}

A theory \HP predicts that a quantum system behaves classically when the theory's  quantum probabilities are high  for histories of the system that exhibit correlations in time summarized by deterministic classical laws. 
The predictions of the classical flight of a tennis ball, the classical orbit of a planet, or the classical behavior of the geometry of the universe are all examples.  

Realistically, classical behavior will not be predicted over infinite stretches of time. The universe behaves classically now but did not behave classically near the big bang nor does it behave classically at  moments of bubble nucleation. More generally,  the theory predicts sets of alternative {\it quasi}classical histories  with high quantum probabilities for classical correlations over stretches of time interrupted by quantum events such as bubble nucleations or the formation and evaporation of black holes \cite{HH15a}.  Most of our large scale observations are of a classical regime of our universe's particular quasiclassical history. Quasiclassical histories are of special interest in false vacuum eternal inflation where bubble nucleations give rise to qualitatively new --- and observationally distinct --- habitable classical regimes of the universe. 

Decoherent sets of alternative coarse-grained histories are called {\it realms}. The histories of a realm are necessarily coarse-grained for decoherence. Beyond that there are realms defined at various levels of further coarse graining.  Coarse grained realms follow some features of the universe and ignore others.  Of particular interest are quasiclassical realms coarse grained by the variables of classical physics including spacetime geometry whose probabilities favor classical evolution. A specific quasiclassical realm which does not ignore the nucleation CDL and HM bubbles in the false vacuum dS background is constructed in the next subsection.

\subsection{Fine and Coarse Grained Bubble Histories}
\label{littleform}

There are different histories of bubble nucleation in the false vacuum at different levels of coarse graining. At a very fine grained level one can consider histories specified by where, when, and what kind of bubble nucleated at various places in the expanding deSitter background. A schematic representation of a particular history of this kind is shown in Figure \ref{bubbles}{\it l}. The quasiclassical ensemble of such histories predicted by \HP does not consist of just one history like that in Fig \ref{bubbles}{\it l}. Rather it is an ensemble of possible ones with bubbles nucleating at different times and different places. This description of this realm is four-dimensional --- diffeomorphism invariant and choice of  slicing independent. 

We shall  assume that the  quantum state  predicts a nonzero probability for the false vacuum background deSitter spacetime in which bubbles can subsequently nucleate. For our purposes it suffices to work in the approximation where we neglect the back reaction of the bubbles on the background spacetime outside the bubble and assume that the bubble collision rate is negligible. On the other hand we must include the fact that the deSitter expansion ends inside bubbles and is replaced by a slow roll inflationary phase towards one of the true vacua. 

In this approximation, a very simple set set of bubble histories can be specified explicitly as follows:   Label the points in the deSitter background by standard coordinates. 
\be
ds^2=-dt^2 + \frac{1}{H^2} \cosh^2(Ht) d\Omega_3^2
\label{desitter}
\ee
where $H^2=3V(0)$ and $d\Omega_3^2$ is the metric on the unit round 3-sphere. 
Consider a discrete sequence of times $t_a=(t_1,t_2, \cdots t_n)$ separated by a time interval $\Delta t$.  At each time divide space up into an exhaustive set of mutually exclusive regions $\Delta_i$, $i=1,2,\cdots$.  A spatial field configuration at any one time is given by specifying for each $\Delta_i$  whether the field is in the range of $A$, $B$ or $F$ where the ranges $A$ and $B$ include the approach to the true vacua. The total number of spatial Hubble volumes $N(t)$ on any one slice increases by a factor $e^{3Ht}$ at least as far as the false vacuum Hubble volumes is concerned. The Hubble volumes where a bubble of true vacuum nucleates decouple from the exponential false vacuum expansion and we do not take in account further quantum events inside those Hubble volumes. 

In general, an exhaustive set of exclusive yes-no alternatives at a moment of time are represented by a set of Heisenberg picture projection operators $\{P_i(t)\}$ satisfying the general conditions\footnote{To be definite we work within decoherent histories formulation if the quantum mechanics of closed systems (DH)\cite{classicDH} extended to semiclassical spacetime \cite{Har93a}, For a tutorial oDH see \cite{Har95c}.}:
\be
\label{projops}
P_{i}(t) P_j(t) = \delta_{ij} P_{i}(t),  \quad  \sum_i P_{i} (t) = I .
\ee
Now we consider specific sets of projections relevant for our problem. 

Pick one volume $i$ at time $t$. The alternatives for the field $A$, $B$ or $F$ can be represented as a set of projections
 $\{\Pss^i_K (t)\}$ where $K$ ranges over the field values. A detailed configuration of the universe at one time is specified by giving $A$, $B$ or $F$ for each spatial volume. We write $\av$ for such a configuration and denote the corresponding set of projections by $\{P_{\av} (t)\}$.  Each $P_{\av}(t)$ is a product of the appropriate  $\Pss^i_K(t)$'s over all volumes $i$.  
 
A history of the universe is a sequence of such spatial field configurations at the series of times. We write for a history $\ab = (\av_1, \av_2, \cdots \av_n)$. Histories are represented by the corresponding chain of projection operators viz 
\be
\label{Cs}
C_{\ab} \equiv P_{\av_n}(t_n) \cdots P_{\av_1}(t_1). 
\ee

The use of a particular slicing in this construction to describe four-dimensional histories is a matter of simplicity and convenience. As mentioned above the histories are defined in a slicing independent way and another slicing should give the same results. Quantum probabilities for histories are defined four-dimensionally and not by singling out one particular slice {\nnf or by specifying a cutoff surface.}

Having understood how a set of histories in false vacuum eternal inflation can be represented in quantum mechanics, we now turn to their probabilities. From the operators \eqref{Cs} branch state vectors $|\Psi_{\ab}\rangle$ can be constructed --- one for each history
\be
\label{branch}
|\Psi_{\ab}\rangle \equiv C_{\ab} |\Psi\rangle . 
\ee
The branch state vector $|\Psi_{\ab}\rangle$ is the quantum amplitude for the history $\ab$ {\nnf in the state $\Psi$}. This set of alternative coarse grained histories decoheres when there is negligible quantum interference between the individual histories as represented by their branch state vectors\footnote{In the classic example of Joos and Zeh \cite{JZ85} a dust grain of millimeter size is in a superposition of positions a millimeter apart deep in space. The characteristic time for the decoherence of alternative position histories of the dust grain by interaction with photons of the cosmic background radiation is about a nanosecond. As mentioned earlier we will assume such mechanisms of decoherence.} viz.
\be
\label{decoherence} 
|\langle \Psi_{\ab} |\Psi_{\bb}\rangle |\ll \sqrt{\langle\Psi_{\ab} |\Psi_{\ab}\rangle \langle \Psi_{\bb}|\Psi_{\bb}\rangle},  \ \ \text{for} \ \  \ab \not= \bb .
\ee
The probabilities $\{p(\ab)\}$  for the histories predicted by \HP are then
\begin{subequations}
\label{probhist}
\be
\label{nrqmprobs}
p(\ab)=||C_{\ab}|\Psi\rangle||^2 ,
\ee
or more explicitly
\be
\label{nrqmexplicit}
 p(\ab) \equiv p(\av_n,\cdots,\av_1) = ||P_{\av_n}(t_n)   \cdots P_{\av_1}(t_1)|\Psi\rangle||^2 .
\ee
\end{subequations}
These are consistent, that is, they satisfy the usual rules of probability theory as a consequence of decoherence\footnote{ For more information on this basic decoherent histories framework in this notation see e.g. \cite{Har93a}.}. 

Quantum mechanics allows us to calculate the probabilities for coarse grained histories by coarse graining the amplitudes for the fine-grained histories.
It is useful to review this in the present context. For simplicity suppose we consider histories $\av$ with three times $t_1$, $t_2$ and $t_3$.  
{\nnf We can calculate the probabilities for coarser grained histories 
by first summing the branch state vectors \eqref{branch}. For instance,}
\be
\label{cgamps}
|\Psi_{\av_3,\av_1}\rangle = \sum_{\av_2} |\Psi_{ \av_3, \av_2, \av_1}\rangle.
\ee
This holds trivially from \eqref{projops} because
\be
\label{nrqm-cg2}
\sum_{\av_2} P_{\av_3}(t_3)P_{\av_2}(t_2)P_{\av_1}(t_1)|\Psi\rangle= P_{\av_3}(t_3)\hcancel[red]{P_{\av_2}(t_2)}P_{\av_1}(t_1)|\Psi\rangle=P_{\av_3}(t_3)P_{\av_1}(t_1)|\Psi\rangle.
\ee
Taking the norm of both sides of \eqref{cgamps} and using the decoherence condition \eqref{decoherence} gives the sum over probabilities,
\be
\label{cgprobs}
p(\av_3,\av_1) = \sum_{\av_2} p( \av_3, \av_2, \av_1)\ .
\ee
Thus coarse graining over amplitudes \eqref{cgamps} is the same as coarse graining probabilities \eqref{cgprobs} as a consequence of decoherence. As \eqref{nrqm-cg2} shows, summing amplitudes over alternatives is the same as ignoring them. It is in fact a special case of sum-over-histories quantum mechanics. 

The description of the universe in terms of the field in an exhaustive set of spatial volumes is not the only possible one. Following the history of the field in a particular spatial volume is an example of a much coarser-grained {\nnf (CG)} description. A {\nnf coarse-grained history of this kind} is specified by giving the {\nnf field value $K$} for the volume at a series of times, $\Kb=(K_n, \cdots, K_1)$. The relevant chain of projections is
\be 
{\sf C}_{\Kb} = \Pss_{K_n}(t_n), \cdots \Pss_{K_1}(t_1)
\label{Khist}
\ee
and the probabilities of these histories
\be
\label{Kprob}
p(\Kb) = ||{\sf C}_{\Kb} |\Psi\rangle||^2
\ee
as in \eqref{nrqmexplicit}. 

These two descriptions are connected. First the operators $C_{\Kb}$ representing the coarser grained histories are sums over the operators $C_{\av}$ representing the finer grained ones. That is because the sum of the projections representing ignored alternatives is unity from \eqref{projops}. {\nnf This means that the probabilities of the fine-grained histories are {\it not needed} to compute probabilities in the CG description. However, if the fine-grained description decoheres - as we assume here - then} the probabilities of the more coarse grained description will be equal to the probabilities of the finer grained one summed over the unfollowed volumes of the mode finer-grained description. 

Bubble nucleation rates could be in principle calculated by using a coarse graining like CG as follows. At a time $t$ pick a particular spatial volume of size $\Delta V$ in the false vacuum. Consider the two time history in which this volume is inside a bubble of type $A$ at a time $\Delta t$ later. From \eqref{probhist} the probability for this history, {\wwf including the local transition} between the false and true vacua is 
\be
\label{transprob}
p(F \rightarrow A) = ||\Pss_{A}(t+\Delta t)\Pss_{F}(t )|\Psi\rangle||^2
\ee
When divided by $\Delta t\Delta V$ {\wwf and in the semiclassical approximation} this gives the nucleation rate $\Gm_A$ for a bubble of kind $A$ per unit four volume multiplied by the probability of $F$ in the quantum state $\Psi$.}

The second description that follows only one spatial volume is much coarser than the first fine-grained (FG) description which follows all spatial volumes\footnote{Two analogous coarse graining were discussed for the box model in Section \ref{YGmodel}.}.
{\nnf We will see shortly that the CG description is appropriate for the calculation of quantum mechanical top-down probabilities for local observations. By contrast the individual histories in FG are qualitatively similar to the description used in TEI.}

The bubble histories of FG {\nnf constructed above} ended at a time $t_n$. Extending them into the future by adding sets of such projections at further times yields an even finer grained set of histories. This risks losing decoherence because there are more and more conditions of the kind \eqref{decoherence} {\nnf that} have to be satisfied. A very simple example\footnote{Stressed to JH often by C. Jess Reidel.} of this \cite{GH90} occurs when the Hilbert space has a finite dimension $N$ either actually or effectively. Then there can be at most $N$ mutually orthogonal branch state vectors $|\Psi_{\ab} \rangle$. A decoherent set can thus have no more than $N$ histories. A set with this number is said to be full \cite{GH90}. Adding one more set of alternatives at one more time to a full set will make the result fail to decohere. {\yf Fine-grained histories describing bubble nucleations extending into the future can be described, but they may not have consistent probabilities.} A probabilistic description of the far future may not be possible without sacrificing (coarse graining) alternatives in the present.
 
But regardless of the properties of these very grained sets of histories they are far more fine-grained than necessary to predict probabilities fore our observations. It is to this we now turn.

\subsection{Top-Down Probabilities for Observations}
\label{cgobs}

{\nnf Top-down probabilities for our observations are conditional probabilities $p({\cal O} \vert \Daoi)$ for different values of observables ${\cal O}$} conditioned on the existence of at least one instance of an observing system described by data $D$ like ours, somewhere in the universe where ${\cal O}$ takes  the given value. {\yf That is all we know from our own observations.} 

To calculate probabilities conditioned on $\Daoi$ we need a model of the existence of an observer. One was sketched in the Introduction. {\nnf In this, an observer like us is described as a physical quantum system within the universe, which} exists in any Hubble volume on the reheating surface of any bubble with a probability $p_E(D)$ --- a very, very small number. In situation where there are $N_h$ relevant Hubble volumes the probability that there is at least one instance of $D$ is 
\be
\label{oneD}
p(\Daoi)=1-(1-p_E(D))^{N_h} \ .
\ee
This is near unity when the number of Hubble volumes is large,  $N_h \gg 1/p_E(D)$ and negligibly small when $N_h \ll 1/p_E(D)$. Bubbles have an infinite number of Hubble volumes on their reheating surfaces and non-eternally inflating universes have a finite number. Thus \eqref{oneD} selects for eternally inflating histories with bubbles, in which the top-down condition on $\Daoi$ is moot\footnote{{\nnf This is a generalisation of a similar result derived in \cite{HH09} in the context of slow roll eternal inflation.} Readers familiar with anthropic reasoning (e.g.~\cite{Tegetal06}) will recognize the top-down factor \eqref{oneD} {\nnf in the $N_h \ll 1/p_E(D)$ regime} as the anthropic selection factor - {\nnf involving a volume weighting $N_h$ and a factor $(p_E)$ plausibly proportional to the fraction of matter in the form of galaxies.} It {\yf might seem that} anthropic selection is impossible in large universes because \eqref{oneD} approaches unity as $N_h$ becomes large.  However \eqref{oneD} is based on the assumption that we are typical of all the instances of our total data $D$. Anthropic selection is possible in large universes {\nnf by conditioning on a part of $D$ only, and including the rest of $D$ as part of the observable ${\cal O}$ to be predicted \cite{HH13}. Top-down probabilities of this kind were computed in \cite{HH13} in the NBWF} to predict the observed value of the cosmological constant and the amplitude of the primordial density fluctuations {\wwf in a toy model landscape.}}.

As we have emphasized before, our observations of the universe are highly coarse grained being restricted to one Hubble volume, inside one bubble, somewhere in the vast inflating background spacetime. A coarse graining  that follows what happens inside our bubble and ignores what goes on outside is therefore adequate for predicting the probabilities for our observations. Assuming the probability of bubble collisions is negligible, our observations tell us simply that we are either in a bubble of type $A$ or $B$. There are then two {\nnf remaining} coarse-grained histories that are relevant, each {\nnf following} only one bubble (ours) but distinguished by whether that is of type $A$ or $B$. 

\begin{figure}[t]
\includegraphics[width=3in]{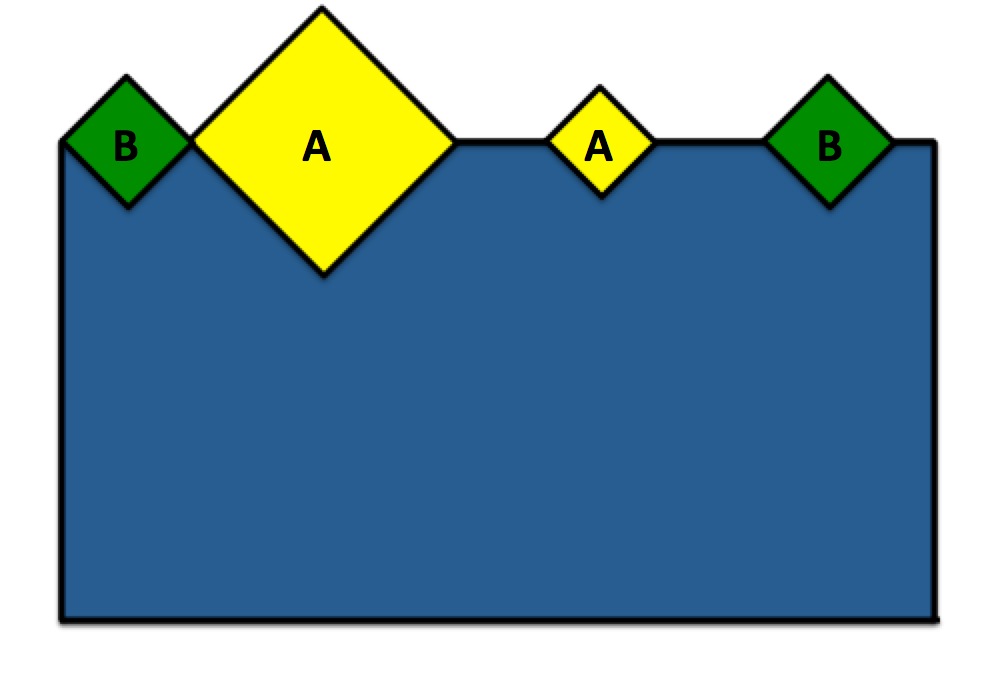}\hfill 
\includegraphics[width=3in]{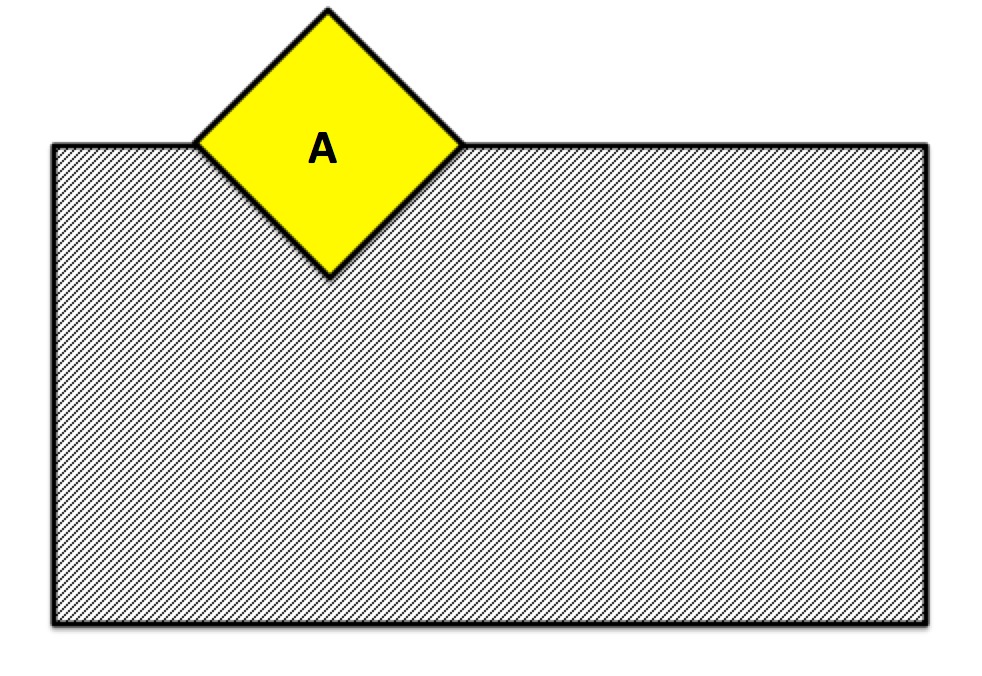}\hfill 
\caption{{\it Left:} Conformal representation of one particular eternally inflating history consisting of a specific configuration of two kinds of bubble universes in a false vacuum de Sitter background. The false vacuum F is indicated in blue, regions inside bubbles of type A are in yellow, and regions inside bubbles of type B are in green. A quantum state of the universe $\Psi$ does not predict one particular eternally inflating history but an ensemble of possible ones.\\ 
{\it Right:} Conformal representation of a coarse-grained history in which the universe inside our bubble is evolving towards the true vacuum A but in which the configuration outside is not specified -- represented by gray. Coarse grained histories of this kind are relevant for the prediction of probabilities for local observations {\wwf in quantum cosmology}.}
\label{bubbles}
\end{figure}

{\nnf We emphasize that this does not mean there are no bubbles outside ours.} For each spatial volume outside ours the probabilities for the alternatives $A$, $B$, and $F$ are summed over (coarse-grained) giving a factor of unity for each volume. {\nnf The histories in the CG description thus provide the total amplitude for a bubble of a certain kind.}
The coarse graining {\it ignores }what goes on outside our bubble. It does not specify it in any particular way. This does not mean that the complex mosaic of bubbles that is the focus of TEI does not exist, but it requires a finer grained set of histories like FG to describe it. {\nnf This is illustrated in Fig. \ref{bubbles}, {\qqf whose similarity with Figure \ref{YGfig} we hope is evident.}}

{\nnf But} how can we evaluate the probabilities for the coarse-grained histories that follow only one bubble when we don't know where our bubble is in spacetime? The answer is that it doesn't make any difference where it is because of the symmetries of deSitter space\footnote{Indeed some notion of symmetry is necessary to make sense of any large or infinite universe. Otherwise anything could happen somewhere.}. In particular the nucleation rates $\Gm_A$ and $\Gm_B$ in \eqref{transprob} are the same everywhere in spacetime {\nnf -- all Hubble volumes are probabilistically equivalent in the background. Hence we can simply choose a location.} Evidently the top-down probability $p({\cal O} \vert \Daoi)$ for a history {\nnf following} one bubble of type $A$ will thus be proportional to $\Gm_A$, and similarly for type $B$. For the probabilities $p(WOA)$ and $p(WOB)$ that we observe {\nnf the properties} of $A$ or $B$ respectively we have then
\be
\label{ans}
\frac{p(WOA)}{p(WOB)} \equiv \frac{p(A \vert \Daoi)}{p(B \vert \Daoi)} = \frac{p(F \rightarrow A)}{p(F \rightarrow B)} = \frac{\Gm_A}{\Gm_B}
\ee
{\nnf where the last equality follows from the fact that both bubbles nucleate in the same false vacuum background, rendering its amplitude in the state $\Psi$ irrelevant. {\wwf We will see in the next section this is a consequence of the simplicity of the potential. In more general landscape potentials the state has an important effect on the top-down probabilities.}

Thus by an appropriate choice of coarse graining we have succeeded in deriving the probabilities for our local observations in this simple model without introducing a measure other than that provided by the quantum state {\nnf and without ever even talking about the fine-grained histories describing a mosaic of bubble universes. As an illustration we now evaluate \eqref{ans} explicitly in the no-boundary state.}

\section{Top Down Predictions of the NBWF}
\label{probnucl}

{\nnf This section discusses more explicitly the top-down predictions of the no-boundary state, first in the model potential in Figure \ref{pot} and then in more general landscape potentials.}

\subsection{Bubble Nucleation Probabilities}
\label{bubnuc}

First we exhibit the {\yf nucleation rates ---} the probabilities per Hubble time and Hubble volume $\Gm_A$ and $\Gm_B$ {\nnf defined in \eqref{transprob}} for a bubble of type $A$ or $B$ respectively to form in the false vacuum $F$. These are essentially specified by the decay rates under or over the barriers separating the true vacua $A$ and $B$ from the false vacuum $F$. 
{\nnf The transition probabilities per Hubble time and Hubble volume have been evaluated semiclassically long ago by Hawking and Moss \cite{HM} and} by Coleman and DeLucia \cite{CDL}. Their result is\footnote{We retain only the exponential factors in the formulae.} 
\be
\Gm_K \propto  e^{-2I_K +2I_{FV}}
\label{nucleate}
\ee
where $I_{FV}$ is the Euclidean action of the false vacuum, $K$ is $A$ or $B$, and $I_K$ is the Euclidean action of the dominant instanton that describes the decay of the false vacuum to the true vacua $A$ or $B$.  

{\nnf When the barrier is large and sharp,  decay proceeds by tunnelling through the barrier as described by a Coleman-DeLuccia instanton \cite{CDL}. Otherwise the decay of the false vacuum occurs via thermal fluctuations over the barrier described by a Hawking-Moss instanton.} Both instantons are regular, compact Euclidean solutions of Einstein equation of the following form,
\be
ds^2=d\tau^2 + a^2(\tau)d\Omega_3^2
\label{CDL}
\ee
{\nnf together with a scalar field profile $\phi(\tau)$.}
The Euclidean action evaluated on an instanton solution of the form \eqref{CDL} is given by 
\be
I[\phi] = \int d^4 x \sqrt{g} V(\phi) = -2\pi^2 \int_0^{\tau_f} d\tau a^3 (\tau) V(\phi(\tau))
\label{CDLaction}
\ee
where $a(\tau_f) = a(0)=0$. The simplest solutions are the $O(5)$-invariant four-spheres in which the field is constant at one of the extrema of $V$ and $a(\tau) = H^{-1} \sin H\tau$. The false vacuum is an example in which $\phi=0$ everywhere. Its action \eqref{CDLaction} is
\be
\label{Fact}
I_{FV} = -\frac{24\pi^2}{V(0)} \ .
\ee
The model potential sketched in Fig. \ref{pot} admits two further four-sphere instanton solutions associated with the maxima of $V$ {\nnf on either side of the false vacuum, in which $\phi(\tau)=\phi_{max,K}$ where $K$ labels the barrier leading to $A$ or $B$.} These are known as Hawking-Moss (HM) instantons and their actions are 
\be
I_{HM} = -\frac{24\pi^2}{V(\phi_{max,K})} \ .
\ee
Even though the field is everywhere constant in the HM instanton, small fluctuations will eventually cause it to roll down to the neighboring true vacuum thereby leading to the decay of the false vacuum. The action of the HM instanton gives the decay rate of the false vacuum through this process.

{\nnf A negative mode analysis shows} that the HM instanton provides the dominant decay channel for relatively broad barriers {\nnf \cite{Gratton00}. Specifically, if $-4 < V_{,\phi \phi}/H^2 (\phi_{max,K}) <0$ then the HM instanton has precisely one negative mode. Hence for barriers of this kind one has}
\be
\label{nucleate}
\Gm_K  \propto  \exp\left(\frac{24\pi^2}{V(\phi_{max,K})}-\frac{24\pi^2}{V(0)} \right) \ .
\ee
The exponent is simply equal to  the difference in entropy of both de Sitter backgrounds. 

{\nnf By contrast, at maxima where  $V_{,\phi \phi}/H^2 (\phi_{max,K}) \leq -4$ the HM instanton has five additional negative modes \cite{Gratton00}.} These are the perturbative indication of the existence of a lower-action non-perturbative solution. This is the Coleman-De Luccia (CDL)  instanton, in which the field {\nnf profile $\phi (\tau)$} straddles the maximum. If $V_{,\phi \phi}/H^2 (\phi_{max,K}) <-4$ the CDL instanton has lower action and precisely one negative mode.

CDL instantons are slightly more complicated solutions of the form \eqref{CDL} in which the field $\phi (\tau)$ varies across the instanton from an initial value $\phi_0$ near the false vacuum to a final value $\phi_{f,K}$ on the other side of the barrier near the true vacuum $K$. In the limit $V_{,\phi \phi}/H^2 (\phi_{max,K}) \rightarrow-4$ the CDL solution tends to HM and so does its action. By contrast, in the limit where the barrier is narrow and sharp and hence $\vert \phi_{f,K} - \phi_0 \vert \ll 1$, gravitational effects become unimportant and the decay rate \eqref{nucleate} implied by the CDL action tends to the well-known thin-wall result \cite{CDL},
\be
\Gm_K \propto  \exp\left(-\frac{27\pi^2 T^4_K}{2 V(0)^3} \right)
\label{CDLthin}
\ee
where $T_K$ is the tension of the narrow barrier separating $F$ from the true vacuum $K$, 
\be
T_K = \int_{\phi_0}^{\phi_{f,K}} d\phi \sqrt{2V(\phi)}.
\ee
{\nnf To summarise, the transition probabilities \eqref{transprob} can easily be computed in the semiclassical approximation, and they depend on the height and the shape of the barriers separating $F$ from $A$ and $B$.}

\subsection{CG Probabilities from the NBWF}
\label{CDLfNBWF}

{\nnf Using this more explicit form of the nucleation probabilities we now return to the probabilities $p(F \rightarrow K)$ of the CG histories following a single bubble with branch state vectors $|\Psi_{K}\rangle \equiv C_{K} |\Psi\rangle $. These probabilities greatly simplify in the no-boundary state, in which case they can be derived directly from the wave function. This is because the no-boundary weighting of the false vacuum \eqref{bu} precisely compensates for the false vacuum factor in the tunneling rate $\Gm _K$ given by \eqref{nucleate}. Indeed the standard analysis of the nucleation probabilities assumes the vacuum state initially. Specifically, this means
\be
p(F \rightarrow K)  \propto  e^{-2I_K}
\ee
where $I_K$ is the action of the dominant instanton mediating the transition from $F$ to the true vacuum $K$ {\it without} the false vacuum action \eqref{Fact} subtracted. The top-down probabilities \eqref{ans} that we observe {\nnf the properties} of $A$ or $B$ then become} 
\be
\label{Pgen}
\frac{p(WOA)}{p(WOB)} =  e^{2I_B - 2I_A} \ .
\ee
{\nnf In fact, as regular compact solutions of the Euclidean field equations the CDL and HM instantons are valid saddle points of the NBWF. In this interpretation \cite{Gratton00}}. CDL instantons specify the no boundary amplitude of a coarse-grained history which follows a single bubble of true vacuum that expands in a false vacuum background. The nucleation point of the bubble is located {\nnf at the throat of the deSitter background in the CDL instanton but, as we have seen, all bubble locations are equivalent. Hence the CDL instanton provides an elegant unified description of the {\it coarse-grained} histories in the NBWF consisting of} the creation of a false vacuum background together with a bubble of true vacuum. This is appealing because the no-boundary condition of regularity also determines the state of the fluctuations inside the bubbles which otherwise requires further assumptions.

\subsection{Landscape potentials}

Finally we turn to top-down predictions in more general potentials. In \cite{Her13} we considered a landscape potential without false vacua but with a range of different inflationary slow-roll patches including power law slopes and plateau-like patches. We showed that in a landscape of this kind, top-down probabilities in the no-boundary state favour histories that emerge from regions of eternal inflation lying at low values of the potential. This selects low plateau-like patches in this landscape and leads to the prediction that we should observe a CMB pattern of fluctuations with statistical features characteristic of plateau-like potentials. 

Our results make it straightforward to generalize the analysis of \cite{Her13} to landscape potentials with false vacua. False vacua constitute a qualitatively new class of eternal inflation patches in the landscape. The exit from eternal inflation is mediated either by the CDL or HM instantons discussed above. Together with the no-boundary amplitudes of the false vacua this specifies the contribution of each false vacuum to top-down probabilities for observations in the NBWF, given by \eqref{Pgen}. The lowest action instanton dominates the decay of a given false vacuum. But the result \eqref{Pgen} also implies a relative weighting of different false vacua specified by the Euclidean action associated with the creation of the false vacuum background itself. False vacua at higher values of the potential are strongly suppressed relative to false vacua at lower values in the NBWF. Therefore the central result of \cite{Her13} that TD probabilities in the NBWF are dominated by the lowest exit from eternal inflation generalises to false vacua models in the following form: {\nnf In a landscape of false vacua in the no-boundary state we predict our local universe emerged from the dominant decay channel of the lowest energy false vacuum}.

{\nnf Throughout this paper it has been} assumed that the bubble nucleation rates were small enough that bubble collisions are negligible. Then our Hubble volume must be located on the reheating surface of either an $A$-type bubble, or a $B$-type bubble. {\nnf In more general landscape potentials one expects there will be false vacua with 
high nucleation rates for which collisions between bubbles become more probable.} A collision of two bubbles is in effect a new kind of bubble leading to different alternatives for observation (see e.g. \cite{Kle11,Kle12,Johnson2015}). To predict these new observations a finer grained set of histories is required that follows not just the inside of a single bubble but also a spacetime region around it. Top-down probabilities are more difficult to compute but still possible. In this way one can imagine systematically fine-graining {\nnf the history of the universe} until further effects on our local observations become negligible.

\section{Conclusion: Classical versus Quantum Measures}
\label{compare}

We have shown that top-down probabilities for our local observations in a universe undergoing false vacuum eternal inflation can be obtained from the basic principles of quantum mechanics and quantum cosmology.
To conclude we compare and contrast our framework for prediction based on quantum cosmology (QCEI) with the traditional approach to the measure problem in eternal inflation (TEI).

TEI and QCEI share the same objectives. Both aim to predict top-down probabilities for the  local observations of observers like us. But TEI is an essentially classical framework in which one starts with a fine-grained description of one of the universe's histories. TEI computes probabilities for observations by counting the number of Hubble volumes (or bubbles) in a typical fine-grained history of the universe where observables take different values. This requires a prescription for regulating infinities, because a fine-grained description typically follows an infinite number of bubbles, each of which is itself infinite. This prescription, which specifies a measure in TEI, supplements the basic theory. It consists e.g.~of specifying a spacelike three-surface beyond which one no longer counts instances of observations. 

It is well known, and hardly surprising, that the resulting predictions are highly regulator- dependent. However, certain choices of TEI measure do yield predictions which are not very different from what we find, at least for observables that are local in time. In particular, measures have been put forward in which the numbers of observations are governed by the nucleation rates of different kinds of bubbles multiplied by the number density of observers, resonating with our findings for the probabilities of a certain class of observables \cite{HH13}. On the other hand the same measures also make predictions that are in conflict with basic physical intuition \cite{EOT}. {\wwf Taken} together with the ambiguities {\wwf mentioned} above this means the TEI approach to define a measure remains deeply unsatisfactory.

By contrast we have shown how in QCEI probabilities for our coarse-grained observations can be calculated simply and directly, without using a fine-grained description, and without a posited probabilistic measure beyond that supplied by the quantum state. The key analogies and differences between the quantum framework put forward here and the traditional measure program can be summarized as follows:

\begin{itemize}

\item{{\it Quantum State of the Universe:} TEI assumes one eternally inflating false vacuum history in which bubbles nucleate and collide. This fine-grained history serves as a background for the calculation of probabilities. 
QCEI instead derives an ensemble of background histories of the universe, with their probabilities, from a quantum state of the universe. This paper has shown that the dominant decay channels of the lowest false vacua provide the largest contributions to top-down probabilities in the no-boundary state. By contrast TEI does not favour low false vacuum backgrounds over others, because the existence of a classical background is assumed.}

\item{{\it Models of Observers:}  Not surprisingly a model of observers is necessary for predicting the probabilities of observations. QCEI, as discussed here, assumes that observers are quantum systems within the universe and not somehow outside it. {\wwf We have adopted} a very simplified model in which observers have evolved with a very small probability in any Hubble volume or have not evolved (e.g.~in the false vacuum). Top-down probabilities are conditional probabilities for different values of observables given the observer/observational situation. The small probability for the observational situation to exist in any Hubble volume has an important effect on the top-down predictions: First, it suppresses the contribution of non-eternally inflating histories to top-down probabilities. Second, it means the condition on the observational situation becomes trivial in eternally inflating bubble histories where the observational situation is certain to exist. While this a very crude model of observers, it is more realistic than the models used in TEI where observers are treated classically, and assumed to exist wherever possible.}

\item{{\it Typicality:}  TEI and QCEI as presented here make use of a similar typicality assumption\footnote{However the notion of typicality is implemented differently. TEI specifies a reference class to define what is meant by an observer. All instances of observers defined this way are treated equally. By contrast in QCEI our particular observational situation is modeled as a subsystem, in terms of a set of data $D$. All instances of this data are then treated equally.}: We are equally likely to be any of the instances of our observing situation in the universe. It is important to stress that there is not a shred of observational evidence one way or the other for this assumption. However it is both simple and natural in the absence of evidence to the contrary. Different assumptions are possible and can be implemented  and tested \cite{HS07}.}

\item{{\it Symmetries:} The eternally inflating deSitter backgrounds that are assumed in TEI, and dominate the classical ensemble in QCEI, have all the symmetries of deSitter space. This means in particular that the nucleation probabilities of bubbles of various kinds are everywhere the same --- all Hubble volumes are probabilistically equivalent. This symmetry plays a significant role in the QCEI calculations of this paper. By contrast, this symmetry is broken in any particular fine-grained quasiclassical history that is the starting point for the calculation of probabilities in TEI. Different locations are no longer equivalent -- they are either in the false vacuum or in one of the different kinds of bubbles -- {\wwf leading one to to take into account observations in all of them}.}

\item{{\it Coarse Graining:}  Our observations of the large scale universe are very coarse-grained. QCEI uses coarse grainings consistent with the quantum mechanical symmetries to directly calculate the probabilities for our observations in a very large universe. The `measures' posited in TEI are a kind of coarse graining because they typically ignore events to the future of some space like surface in the background spacetime\footnote{Evidently one cannot adopt a coarse-graining that follows a single bubble only in TEI, because then there would be no bubbles left to count. A sufficiently fine-grained description is needed in TEI to accommodate the different possible observations.}. However such measures break the symmetries that are basic to QCEI. The resulting difficulties are well documented in the extensive TEI literature (see e.g.~\cite{ProblemsTEI}).}

\item{{\it Source of probabilities:} TEI is an essentially classical framework for prediction whereas QCEI is based on quantum cosmology. This also means that the source of the probabilities that are computed is different. The probabilities derived by QCEI are quantum probabilities --- squares of amplitudes as in Born's rule. In contrast the probabilities in TEI are classical representing ignorance  --- ``we don't know where we are.''  The latter are therefore calculated by counting instances\footnote{Connections between Born's rule and counting have been discussed by \cite{AT10,BS11,Vil13}.} of our observational situation in a typical {\zf realization/instantiation} of the complex fine-grained global structure of the universe using a particular cutoff procedure that breaks the symmetries.
Counting an infinite number of things is certainly difficult and most likely ambiguous. Instead we have shown that the quantum mechanical symmetries make it possible to coarse grain to the finite number of possibilities for our observations directly by summing amplitudes.}

\end{itemize}

This assessment does not mean there is no measure problem in quantum cosmology. There is one, but it is not the problem of what regulates divergences in specific calculations. It is rather the question of what is the quantum state of the universe.

\vskip .2in


\noindent{\bf Acknowledgments:} We thank the participants of the KITP Quantum Gravity program for stimulating discussions. TH thanks the KITP and the Physics Department at UCSB for their hospitality. The work of JH was supported in part by the US NSF grants PHY12-05500 and PHY15-04541. He thanks the theoretical physics group at KU Leuven and the Solvay Institutes for support during visits there. The work of TH is supported by the European Research Council, grant no. ERC-2013-CoG 616732 HoloQosmos, and by the Belgian National Science Foundation (FWO).

\appendix
\section{ Better Box Models}
\label{bbm} 

 Section \ref{YGmodel}'s box model assumed unit probabilities $p(\Eo|A)$ and  $p(\Eo|B)$  that there is at least one observer like us somewhere inside any bubble \eqref{pE1}. As we will explain better below, that is a correct assumption if there are an infinite number of Hubble volumes on the reheating surfaces of both kinds of bubbles. But were the surfaces not infinite,  or the probabilities \eqref{pE1} not unity, and/or different from each other, the probabilities that we observe $CMB_A$ or $CMB_B$ can be significantly changed. The aim of this appendix is to illustrate how this works in a simple model. 
 
 The model is the same as in Section \ref{YGmodel} with one extension:  Each box of type $A$ is assumed to contain  $N_h^A$ Hubble volumes  that model the volumes on type $A$'s reheating surface.  We assume a probability $p_{E}^A$ that an observer like us evolves in any of these Hubble volumes. This probability models all of the accidents of galactic, planetary and biological evolution and is therefore a very small\footnote{It is a number that is well beyond our ability to compute in present day physics as illustrated by the effort in \cite{Koo11}. See \cite{HH09,HH13}  for further discussion.} number! The same kind of assumptions are made for bubbles of type $B$ with the corresponding quantities denoted by$N_h^B$ and $p_{E}^B$.   Note that $p_E^A$ is the probability for an observer to have evolved in any {\it Hubble volume} on  the reheating surface of {a bubble of kind $A$}, whereas $p(\Eo|A)$ is the probability that at least one observer evolves somewhere on the reheating surface of {\it a bubble} of kind $A$. We derive the  connection in \eqref{td}.
 
 The fine-grained histories for this model describe whether each of the $N$ boxes  is of type $A$, $B$, or $F$ and whether at least one observer exists in each Hubble volume inside each  box or not.   Our observations are much more coarse grained --- confined to one Hubble volume in one box of kind $A$ or $B$.   Probabilities relevant for the observations by our particular instance of an observing system are necessarily conditioned on a description of our observational situation.  In the context of this simple model our observational situation is  described by the fact that there exists at least one observer ($\Eo$)  --- us--- in some Hubble volume in some bubble. 

 The top-down probabilities relevant for our observations are  therefore the conditional probabilities given this data that we are in a box of kind $A$ or $B$. We write these  $\pfp(A|\Eo)$ and $\pfp(B|\Eo)$. These conditional probabilities can be deduced from joint ones  \begin{subequations}
  \label{cj}
 \begin{align}
 \label{cj1}
 \pfp(\Eo,A) &= p_A p(\Eo|A) ,   \\ 
  \pfp(\Eo,B) &= p_B p(\Eo|B). \label{cj2}
\end{align}
\end{subequations}

The probababilty $p(\Eo|A)$ that there is at least one observer in one of the $N_h^A$ Hubble volumes on the reheating surface on a bubble of type $A$ is $1$ minus the probability that there are no observers in any Hubble volume. The  latter is the probability $1-p_E^A$ that there is no observer in the Hubble volume raised to the number of Hubble volumes. Thus we have
 \be
 \label{td}
 p(\Eo |A) = 1 -(1-p^A_{E})^{N^A_h} ,
 \ee
 and similarly for type $B$. Thus for the top-down probabilities we have 
 \begin{subequations}
\label{probs}
\begin{align} 
 p(WOA)\equiv\pfp(A|\Eo) &= \frac{p_A p(\Eo,A)}{p_A p(\Eo,A)+p_Bp(\Eo,B)} ,  \label {probs1} \\
  p(WOB)\equiv\pfp(B|\Eo) &= \frac{p_B p(\Eo,B)}{p_A p(\Eo,A)+p_Bp(\Eo,B)} .  \label {probs2}
\end{align}
\end{subequations}
This is the general result but  two limiting cases are of interest:
 
 {\it  Infinite reheating surfaces in all bubbles:}   This the usual assumption following from the Coleman-DeLuccia  analysis.  In this case the probability that there exists a copy of us somewhere is unity as shown explicitly by \eqref{td} when $N_h^A$ and $N_h^B$ are infinite. Then $p(\Eo|A)=p(\Eo|B)=1$ as was assumed in Section \ref{YGmodel}'s box model and we have 
 \begin{subequations}
\label{infH}
\begin{align} 
 p(WOA)\equiv\pfp(A|\Eo) &= \frac{p_A}{p_A +p_B} ,  \label {inHA} \\
p(WOB)\equiv\pfp(A|\Eo) &= \frac{p_B }{p_A +p_B}  \label{infHB} .
\end{align}
\end{subequations}
This is the result quoted in \eqref{obs} and \eqref{obs1}. More precisely it holds when $N_h^A p_E^A \gg 1$ and $N_h^B p_E^B \gg 1$.

In the opposite limit when the number of Hubble volumes in both kinds of  bubble is small  in the sense that $N_h^A p_E^A \ll 1$ and $N_h^B p_E^B \ll 1$ we have
 \begin{subequations}
 \label{rare}
 \begin{align}
p(WOA)\equiv\pfp(A|\Eo) &= \frac{p_A N_h^Ap_E^A }{p_A N_h^A p_E^A+p_B N_h^B p_E^B}  \label{rare1}, \\ 
p(WOB)\equiv \pfp(B|\Eo) &= \frac{p_B N_h^B p^B_E }{p_A N_h^A  p_E^A+p_B N_h^B p_E^B} .\label{rare2}
\end{align}
\end{subequations}
This is the limit that is easiest to understand.  Suppose there was only one Hubble volume in each kind bubble.  Then the probability that we are in a bubble of kind $A$ is the probability that the bubble occurs multiplied by the probability that we are in it.

Of course there are various further combinations of these cases that can be worked out.


\vskip .5in

\end{document}